\documentclass[pdflatex,sn-mathphys-num]{sn-jnl}

\usepackage{graphicx}%
\usepackage{multirow}%
\usepackage{amsmath,amssymb,amsfonts}%
\usepackage{amsthm}%
\usepackage{mathrsfs}%
\usepackage[title]{appendix}%
\usepackage{xcolor}%
\usepackage{textcomp}%
\usepackage{manyfoot}%
\usepackage{booktabs}%
\usepackage{algorithm}%
\usepackage{algorithmicx}%
\usepackage{algpseudocode}%
\usepackage{listings}%
\usepackage{physics}

\usepackage{dcolumn}
\usepackage{bm}
\usepackage[T1]{fontenc}
\usepackage[utf8]{inputenc}

\DeclareUnicodeCharacter{2212}{-}
\DeclareUnicodeCharacter{03B1}{$\alpha$}
\DeclareUnicodeCharacter{03B2}{$\beta$}
\DeclareUnicodeCharacter{03B3}{$\gamma$}

\begin{document}

\title[Article Title]{Lattice matched heterogeneous nucleation eliminate defective buried interface in halide perovskites}

\author[1*]{\fnm{Paramvir} \sur{Ahlawat}}

\author[2,3,4]{\fnm{Cecilia} \sur{Clementi}}

\author[2]{\fnm{F\'elix} \sur{Musil}}

\author[1]{\fnm{Maria-Andreea} \sur{Filip}}

\affil[1]{\orgdiv{Yusuf Hamied Department of Chemistry}, \orgname{University of Cambridge}, \orgaddress{\street{Lensfield Road}, \city{Cambridge}, \postcode{CB2 1EW}, \country{United Kingdom}}}

\affil[2]{Department of Physics, Freie Universit\"at Berlin, Arnimallee 12, 14195 Berlin, Germany}

\affil[3]{Department of Chemistry, Rice University, Houston, Texas 77005, United States}

\affil[4]{Center for Theoretical Biological Physics, Rice University, Houston, Texas 77005, United States}


\abstract{Metal halide perovskite-based semi-conducting hetero-structures have emerged as promising electronics for solar cells, light-emitting diodes, detectors, and photo-catalysts. Perovskites' efficiency, electronic properties and their long-term stability directly depend on their morphology~\cite{mitzi_transport_1995, mitzi_conducting_1995, kojima_organometal_2009, lee_efficient_2012, kim_lead_2012, burschka_sequential_2013, snaith_anomalous_2014, de_quilettes_impact_2015, park_towards_2016,  udayabhaskararao_nucleation_2017, peng_dotwireplateletcube:_2018, correa-baena_homogenized_2019, jena_halide_2019, kosar_unraveling_2021, akkerman_genesis_2018, kim_strategies_2019, xue_surface_2020, li_critical_2021, dey_state_2021, arora_kinetics_2022, han_roadmap_2022, dong_allinorganic_2022, ma_facet_2022, park_controlled_2023}. Therefore, to manufacture stable and higher efficiency perovskite solar cells and electronics, it is now crucial to understand their micro-structure evolution. In this study, we perform molecular dynamics simulations to investigate the formation of cesium lead bromide perovskite on interfaces. Our simulations reveal that perovskite crystallizes in a heteroepitaxial manner on widely employed oxide interfaces. This could introduce the formation of dislocations, voids and defects in the buried interface, and grain boundaries in the bulk crystal. From simulations, we find that lattice-matched interfaces could enable epitaxial ordered growth of perovskites and may prevent defect formation in the buried interface.}


\newpage



\maketitle

*Correspondence to paramvir.chem@gmail.com

\section{Introduction}\label{sec1}
Cesium lead bromide~\cite{wells_uber_1893} has emerged as a versatile all-inorganic perovskite with applications spanning light-emitting diodes~\cite{kim_metal_2016, lin_perovskite_2018, utzat_coherent_2019, hassan_ligand-engineered_2021}, liquid crystal displays~\cite{helio_nodate, avantama}, photo-detectors~\cite{ stoumpos_crystal_2013}, photo-catalysts~\cite{zhu_lead_2019, dubose_efficacy_2022, dubose_energy_2022}, and perovskite solar cells (PSCs)~\cite{duan_high-purity_2018, li_inorganic_2023, wang_suppressed_2023}. Remarkably, this material exhibits room-temperature macroscopic quantum phenomena, such as super-fluorescence~\cite{SF1, SF2, SF3, SF4, SF5, shcherbakov-wu_persistent_2022}, arising from a quantum-coherent ensemble similar to the Bose-Einstein condensate. This material is actively utilized in the fabrication of certified all-perovskite tandem solar cells achieving solar-to-power conversion efficiencies exceeding 28\%~\cite{duan_cspbbr_2023,lin_all-perovskite_2023, nrel} surpassing the market-leading silicon-based solar cells. Furthermore, in perovskite-silicon tandem configurations, this material has been employed as a thermodynamically stable perovskite among mixtures to develop certified solar cells with efficiencies exceeding 30\%~\cite{chin_interface_2023, de_wolf_tandems_2023, nrel}, surpassing the efficiency of gallium arsenide-based solar cells. In the past decade, several experiments~\cite{koscher_essentially_2017, akkerman_strongly_2017, akkerman_genesis_2018, peng_arm_2019, gualdron-reyes_controlling_2019, wu_highperformance_2020, dey_state_2021} have been conducted to improve the control over the shape, size, composition, and quality of the synthesized cesium lead bromide (CsPbBr\textsubscript{3}) perovskite crystals. These experiments involved diverse methodologies, including solution processing, melt crystallization, and vapor deposition. Protesescu \textit{et al.}~\cite{protesescu_nanocrystals_2015} pioneered a shape-controlled synthesis approach using a hot-injection methodology, manipulating reaction temperatures. Subsequently, ligand-mediated synthesis protocols~\cite{zhang_brightly_2015, li_cspbx_2016} were introduced, incorporating organic molecules during growth. Pan \textit{et al.}~\cite{pan_insight_2016} and Sichert \textit{et al.}~\cite{sichert_quantum_2015} employed variable chain length carboxylic acids and amines to control nanocube and nanoplatelet dimensions. Liu \textit{et al.}~\cite{liu_highly_2017} and Wu \textit{et al.}~\cite{wu_improving_2017} further improved synthesis by introducing branched capping trioctylphosphine molecules, enhancing electronic properties. 

\bigskip
However, solvent and organic molecular additives, while beneficial for shaping, can introduce defects, affecting stability and optoelectronic properties~\cite{barker_defect-assisted_2017, lai_intrinsic_2018, zhang_phase_2019}. Alternatively, CsPbBr\textsubscript{3} can be manufactured through melt processing or vapor deposition, offering advantages like the synthesis of large single crystals~\cite{song_ultralarge_2017, he_high_2018, wang_synthesis_2020, liu_one_2022}. Currently, there are many ongoing efforts at both the academic~\cite{maceiczyk_microfluidic_2017, wu_highly_2019, gualdron-reyes_controlling_2019, shamsi_metal_2019, burlakov_competitive_2020, dey_state_2021, kipkorir_cspbbr_2021, pradhan_why_2021,  baek_mechanochemistry-driven_2022, si_efficient_2017, dong_allinorganic_2022, han_roadmap_2022, das_facets-directed_2022} and the industrial (Quantum Solutions~\cite{quantum_solutions}, Avantama~\cite{avantama}, Peroled~\cite{peroled}, Nanolumi~\cite{nanolumi}, Zhijing Nanotech~\cite{zhintech} and Helio Display Materials~\cite{helio_nodate}) level to improve existing synthesis processes and create new ones by using various additives and optimizing process conditions for enhancing the crystalline microstructure of halide perovskites.

\bigskip
 
To achieve precise control over CsPbBr\textsubscript{3} crystalline morphology, one requires an understanding of its nucleation and growth processes at the atomic scale. Ongoing experimental efforts often heavily rely on trial-and-error engineering approaches, while few groups endeavor to characterize the morphology of the synthesized crystals. In recent studies, Manna and collaborators~\cite{zhang_stable_2022} utilized X-ray crystallography, while Kovalenko and co-workers~\cite{akkerman_controlling_2022} employed \textit{in-situ} optical absorption spectroscopy to examine the size and shape evolution of locally nucleated structures.  However, the spatial and temporal resolution of state-of-the-art experimental techniques is often too limited to resolve nucleation processes accurately. Recently, advancements in \textit{in-situ} transmission electron microscopy (TEM)~\cite{nakamuro_capturing_2021, chevalier_precision_2022} have allowed for atomic-level visualization of phase transitions. Nevertheless, the electron-beam sensitivity of halide perovskites, coupled with their soft nature, presents challenges in comprehending their crystallization mechanism through TEM experiments.  Despite its fundamental importance and critical role in defining device stability and efficiencies, the crystallization mechanism of halide perovskites remains elusive, underlying the need for a more comprehensive understanding of these materials.
\bigskip

Alternatively, all-atom molecular dynamics (MD) simulations~\cite{rein_ten_wolde_numerical_1996, wolde_enhancement_1997, cacciuto_onset_2004, piana_simulating_2005, piana_understanding_2005, valeriani_rate_2005, kawasaki_formation_2010, van_meel_design_2010, anwar_uncovering_2011, lechner_role_2011, chakraborty_evidence_2013, zimmermann_nucleation_2015, espinosa_crystal-fluid_2015, salvalaglio_molecular-dynamics_2015, espinosa_seeding_2016, jungblut_pathways_2016, anderson_predicting_2017, shibuta_heterogeneity_2017, elm_modeling_2020, arjun_unbiased_2019} offer a powerful methodology for describing atomic-level details of nucleation and growth processes~\cite{ahlawat_atomistic_2020, lu_vapor-assisted_2020, ahlawat_combined_2021, ahlawat_crystallization_2023, ahlawat_size_2024}, facilitating the design of more targeted and improved experiments~\cite{ahlawat_combined_2021}. In this study, we conduct all-atom MD simulations to explore perovskite nucleation from a homogeneous mixture of ions on interfaces. 

\section{Results}\label{sec2}

\begin{figure}
  \includegraphics[width=1.0\linewidth]{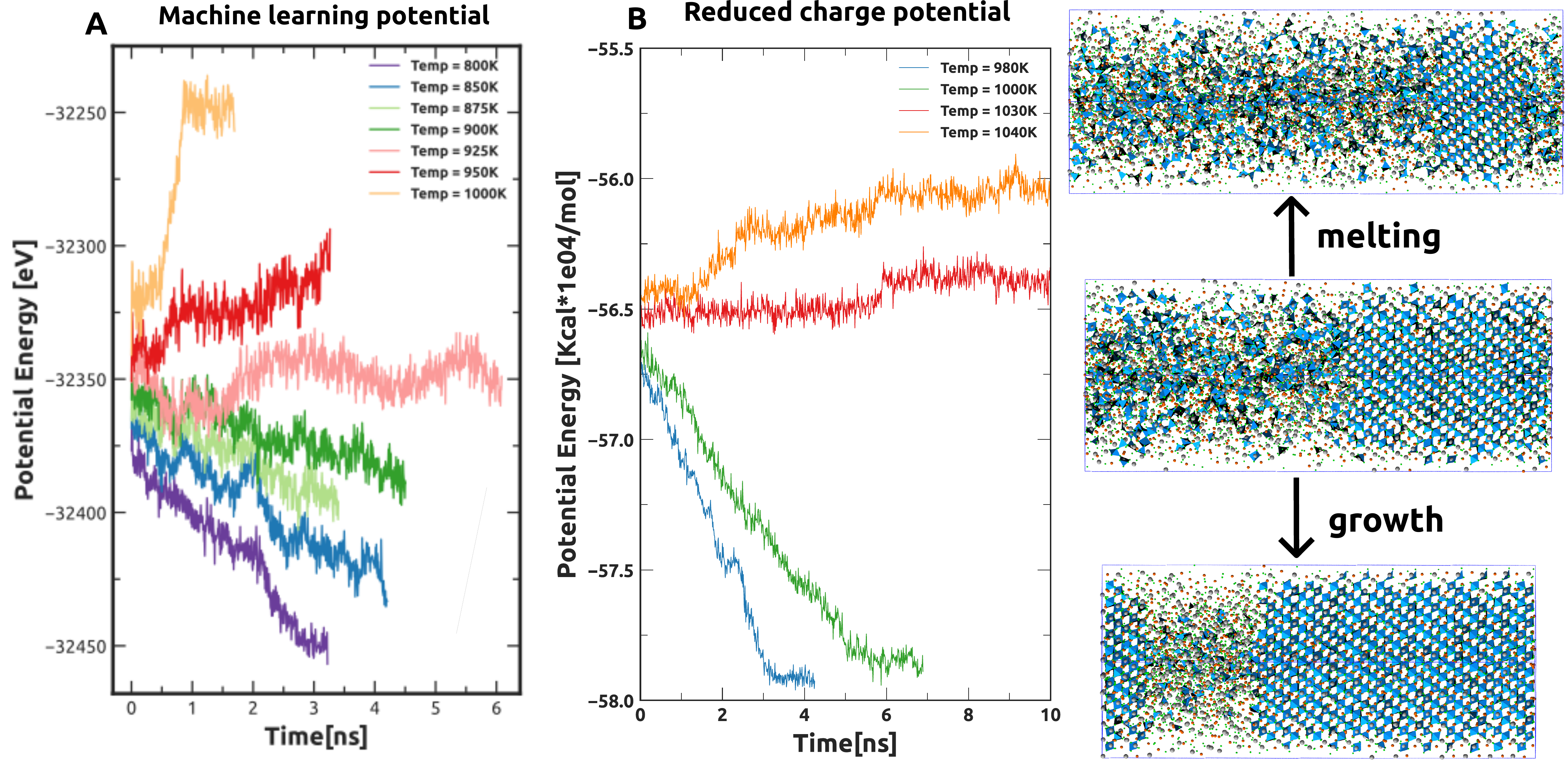}
  \caption{\textbf{Co-existing simulations:} The figure illustrates the changes in potential energy during crystal growth simulations at different temperatures. The panel on the right displays the co-existing configurations and their evolution during melting and growth processes.}
  \label{fig:coexist}
\end{figure}

First, we perform simplified \textit{ab-initio} quantum chemical calculations to develop a scaled-charge~\cite{blazquez_madrid-2019_2022, zeron_force_2019} inter-atomic potential specifically tailored for perovskite, further details are provided in the Methods section. This potential predicts a melting temperature (T\textsubscript{m}) of around 1000K, as determined by co-existing simulations depicted in Figure \ref{fig:coexist}, see Methods section for details. Our predicted T\textsubscript{m} value falls within a 20\% range of the experimentally determined melting temperature (T\textsubscript{expm}) of 860$\pm$40K~\cite{stoumpos_crystal_2013, kanak_melting_2022, zhang_synthesis_2018}. Given the absence of experimental data on the dynamical aspects of perovskite crystallization at the nanosecond timescales, we also conducted a comparative analysis of the crystallization process. This involved comparing the performance of our scaled-charge inter-atomic potential with that of a more expressive machine learning interatomic potential (MLIP) generated by the NEQUIP~\cite{batzner_e3-equivariant_2022} code using r2SCAN+rVV10 level density functional theory (DFT) reference data, as outlined in the Methods section. In Figure \ref{fig:coexist}, our results demonstrate that in both scenarios, perovskite growth and melting events occur at the nanosecond timescale, underscoring the efficacy of our computational models in capturing these dynamic processes.
\bigskip

\subsection*{Homogeneous nucleation}

\begin{figure}
  \includegraphics[width=\linewidth]{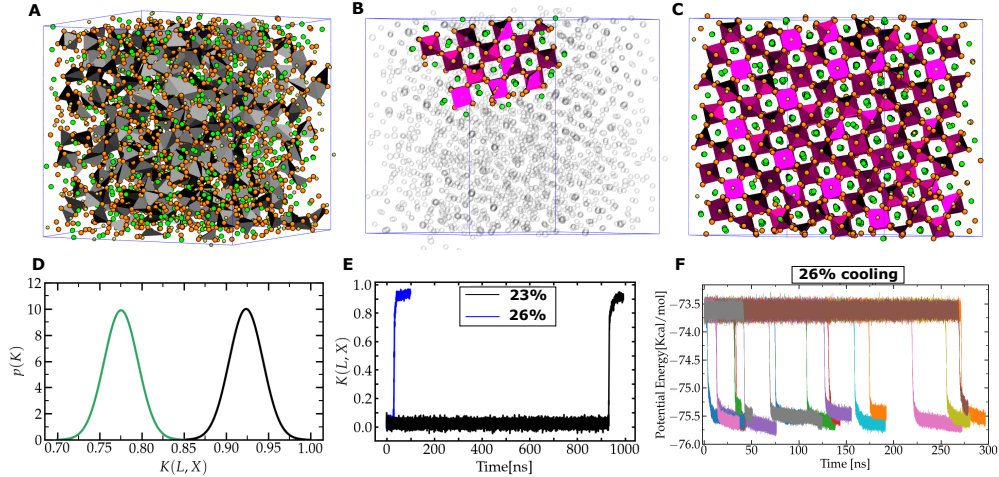}
  \caption{\textbf{Homogeneous nucleation of a perovskite:} (a) shows the initial homogeneous mixtures of ions. Pb-Br complexes are shown as polyhedra with Pb in the center and Br in the corners. Free Cs and Br are shown as green and orange spheres respectively. (a)-(c) shows the evolution of homogeneous nucleation of perovskites at $\sim$ 26$\%$ cooling with (a) being the initial homogeneous mixtures of ions. To guide the eye we show Pb-Br octahedra with separate colors for homogeneous mixtures (a) and crystalline perovskite structures in (b) and (c). (d) shows the probability distribution of SOAP similarity index ($K{(\mathbf{L}, X)}$) for homogeneous melt (green) and perovskite phase (black) of CsPbBr\textsubscript{3}. (e) shows the time evolution of normalized $K{(\mathbf{L}, X)}$ values for the MD trajectory at two different supercooled temperatures. (f) represents the time evolution of the potential energy at $\sim$ 26$\%$ supercooling with variable momenta. All of the above nucleation images are generated with VMD ~\cite{humphrey_vmd:_1996}.}
  \label{fig:homo_nucle}
\end{figure}

We then start to study the process of homogeneous nucleation in perovskite, a critical step in its morphological evolution, using MD simulations with the scaled charge inter-atomic potentials. We initiated our simulations by constructing a system comprising 2560 ions, equivalent to the 512 formula units of CsPbBr\textsubscript{3}, within a periodic cubic box under atmospheric pressure. Under constant temperature-pressure MD simulations, we systematically cooled the systems to temperatures 10\%, 20\%, 25\%, 30\%, and 40\% below the perovskite calculated melting temperature (T\textsubscript{m}). Interestingly, at cooling levels of 30\% and 40\%, we observe the spontaneous emergence of the perovskite structure. Upon further inspecting the atomic level details of the crystallization process, we observe a non-spherical nucleus, already exhibiting macroscopic morphological features characteristic of a faceted crystal structure, as illustrated in Figure \ref{fig:homo_nucle}B and Supplementary Movie SM1. This behavior is similar to the observations of previously studied ionic systems such as NaCl~\cite{valeriani_rate_2005}. It is crucial to note that nucleation is a stochastic process, and our simulations, replicated with different momenta under constant 26\% cooling conditions, exhibit a consistent behavior, as shown in Figure \ref{fig:homo_nucle}F. However, as the degree of cooling decreased below 20\%, no discernible perovskite formation is observed in our extensive $\sim$1$\mu$ second-long of brute force MD simulations. This suggests that nucleation events may become rare events at lower cooling rates, and we may require extensive sampling to understand the dynamics governing perovskite crystallization. 

\bigskip

In our investigation of the crystallization process, we also aim to unravel the intricate structural transitions within multi-species systems such as the one in this study. We introduce a novel collective variable (CV), denoted by $K(L,X)$, that combines a general representation of the atomic neighborhood with the use of target configurations. The \textbf{landmark-SOAP CV} is based on the SOAP kernel~\cite{Bartok2013, Musil2017} where we compare the structural similarity\cite{gallet_structural_2013, Car20218} between handpicked landmark configurations $L$, e.g. perfect perovskite crystals, and a set of configurations $X$, see the Methods section for more details. The SOAP representation of a local neighborhood is general~\cite{Musil2021rev}, so it is tailored to distinguishing between the liquid and perovskite phases of CsPbBr\textsubscript{3} by comparing new configurations with perfect perovskite configurations. Figure \ref{fig:homo_nucle}D shows the probability distributions~\cite{tribello_plumed_2014, bonomi_plumed:_2009} of $K(L, X)$, derived from a combination of molecular dynamics trajectories representing both the perovskite crystal and its molten structure. In particular, a $K(L, X)$ value exceeding 0.85 signifies ions associated with the perovskite structure, thereby facilitating the quantification of the perovskite fraction during the nucleation process. While our current study does not include calculations of nucleation rates, it is important to highlight the potential of $K(L,X)$ in extending our understanding to include such analyses. Specifically, by analyzing mean free passage time data obtained from repeated simulations at various cooling rates, $K(L,X)$ can be extended to calculate nucleation rates in these multi-species systems. Such insights hold promise for elucidating the kinetics underlying complex structural transformations in a diverse range of multi-component materials.

\bigskip

\begin{figure}
  \includegraphics[width=1.0\linewidth]{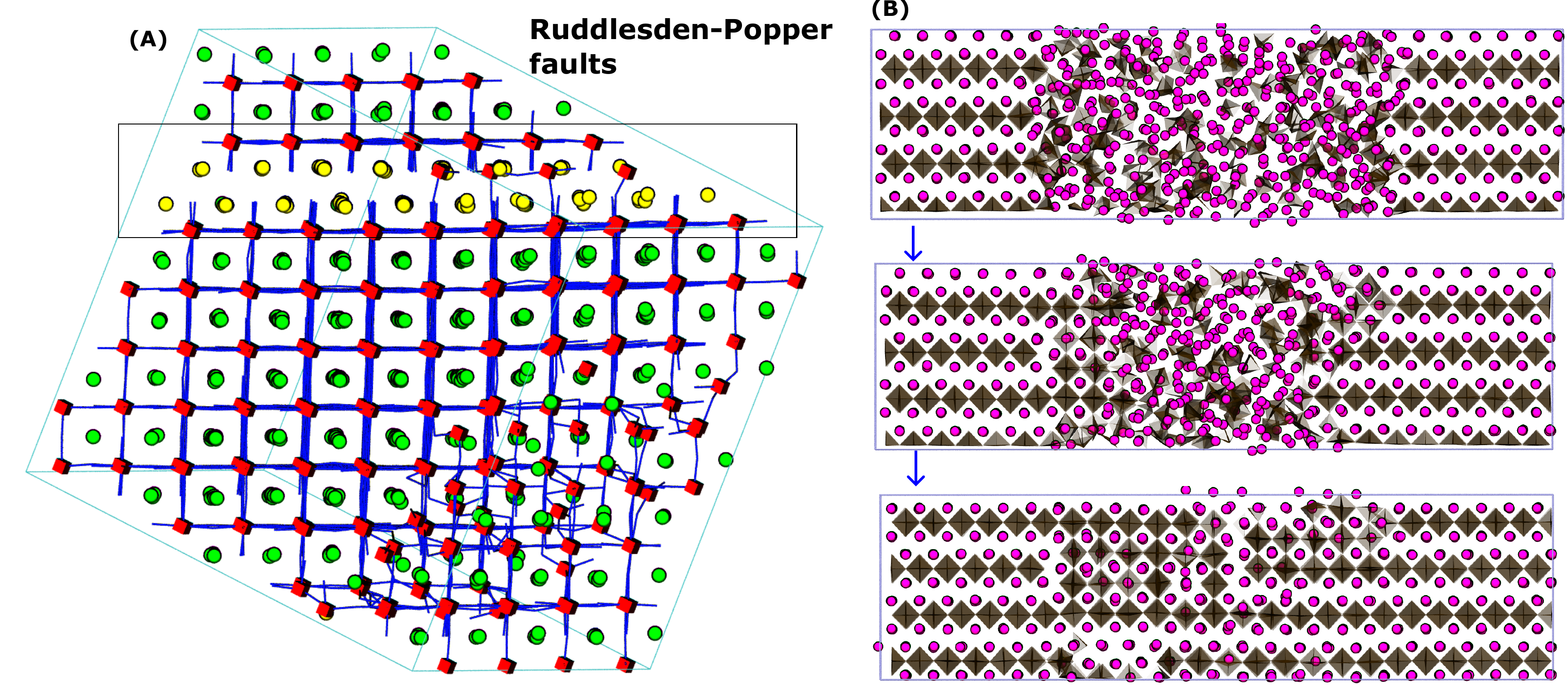}
  \caption{\textbf{Ruddlesden-Popper (RP) faults:} (A) shows RP-faults in a nucleated crystal. (B) illustrates the co-existing simulations of the RP structure alongside a homogeneous mixture of ions.}
  \label{fig:RP}
\end{figure}

For our system, the temporal evolution of $K(L, X)$ presented in Figure \ref{fig:homo_nucle}E unveils a distinctive first-order phase transition during the nucleation of perovskite, devoid of any intermediate meta-stable phases. However, subsequent investigations of post-nucleation crystals reveal the emergence of bulk quasi-two-dimensional Ruddleson-Popper (RP) faults. These planar boundary anti-phase structures are characterized by two consecutive [CsBr] layers between neighboring [Pb-Br] octahedral sites within the perovskite, also highlighted in Figures \ref{fig:RP}A and Supplementary Figure 1. We extracted the RP-faulted structures from our nucleated trajectories and conducted variable-cell-relaxation using DFT. Notably, the structures persist in their RP form, and the DFT-optimized configurations are included in the Supplementary Material. Significantly, the presence of RP-faults in CsPbBr\textsubscript{3} is corroborated by experimental findings~\cite{thind_atomic_2019}, with various atomic-level TEM measurements detecting these structural features in bulk perovskite thin-films. To further investigate the process of RP structures' formation, we carried out co-existing simulations of quasi-2D Cs\textsubscript{2}PbBr\textsubscript{4} seeded structure alongside a homogeneous mixture of ions, as illustrated in Figure \ref{fig:RP}B and and Supplementary Movie SM2. Surprisingly, even in the presence of quasi-2D seeds, we observe the crystallization of mixed quasi-2D and 3D perovskite configurations. Therefore, it is unlikely that these anti-phase RP faults form in bulk during the nucleation of CsPbBr\textsubscript{3} perovskite, suggesting their possible formation at boundaries when two growth fronts interact with each other.

\bigskip

Experimental investigations~\cite{wang_tailoring_2023} have revealed that the density of RP faults in CsPbBr\textsubscript{3} can significantly impact its photoluminescence properties. Moreover, RP-faulted structures are not exclusive to CsPbBr\textsubscript{3}, and they have been observed in various materials. For example, recent TEM experiments~\cite{kim_atomic-scale_2023, shi_premelting_2023} have linked pre-melting phenomena in oxide perovskites to the presence of RP faults. Notably, a study demonstrated a substantial increase in the superconducting temperature of nickelates~\cite{lee_linear--temperature_2023}  upon the elimination of these RP-faulted structures by optimizing their heterogeneous crystallization process. Our simulation findings are in line with experimental measurements across a diverse range of perovskite systems, underscoring the complexity of the crystallization process in multi-species systems, where various condensed phases can emerge. The L-SOAP-CV plays an important role in enabling the differentiation between distinct phases and facilitating a comprehensive atomic-level understanding of the process. Motivated by these exciting observations and successful comparisons with experiments and calculations, we proceed to simulate nucleation on interfaces.
\bigskip

\subsection*{Crystallization on interfaces}
We initiate the process by depositing a homogeneous mixture of ions onto a substrate. We construct simplified systems where Cs, Pb, and Br ions are arranged on the surface of a commonly used anatase-TiO\textsubscript{2} substrate. For this study, we consider two different experimental setups: (A) the crystallization of perovskite on a single flat surface, and (B) the crystallization of perovskite sandwiched between flat surfaces. The latter condition mimics experimental conditions resembling the mesoporous architecture of the substrate, where perovskite may crystallize between large oxide nanoparticles.

\bigskip

Using these two setups, we conduct molecular dynamics (MD) simulations, as outlined in the Methods section. These simulations are performed at a cooling rate of 25\% to prevent rapid perovskite crystallization. Through brute force MD simulations with setup (A), we observe the transformation of the initial ion mixture into the corner-sharing perovskite structure. Nucleation of perovskite initiates from the substrate, and subsequently, the ion mixture fully converts into a large perovskite crystal with accelerated growth. Given the stochastic nature of crystallization, we repeat our simulations five times, consistently observing substrate-induced heterogeneous nucleation of perovskite in all runs. No instances of homogeneous nucleation are observed in the repeated simulations. The complete crystallization process is illustrated in Figures \ref{fig:second}D to H, with an atomic-level view provided in Supplementary Movie SM3.

\bigskip

Crystallization of semiconductors on interfaces typically manifests through three distinct modes: Frank–van der Merwe, Volmer–Weber, and Stranski–Krastanow. Key differences between these modes lie in how the initial crystalline layer nucleates on the interface, as depicted in Figures \ref{fig:second}A, \ref{fig:second}B, and \ref{fig:second}C, respectively. Using the L-SOAP-CV, we characterize our simulations and investigate which of these mechanisms best describes the crystallization of perovskite on oxide substrates. We partition our simulation box into layers parallel to the substrate, with each layer represented by different colors in Figure \ref{fig:second}D. The time evolution of perovskite octahedra in each layer is shown in Figure \ref{fig:second}I.

\bigskip

The perovskite growth extends to the fourth (yellow) layer even before the complete crystallization of the first (blue) layer closest to the substrate. This observation is further illustrated in Figure \ref{fig:second}E, where an island of perovskite has emerged within the simulation box. We note that despite appearing as separate entities, the two islands in Figure \ref{fig:second}E are integral parts of the same crystal, owing to the periodic nature of our simulation box. This means that our analysis identifies the Volmer-Weber (VW) mechanism (illustrated in Figure~\ref{fig:second}B) as governing the perovskite crystallization. This mechanism is also reminiscent of highly mismatched growth in heterostructures as it entails island-type growth on the substrate without the formation of a wetting layer. The significant lattice mismatch between the unit cells of anatase-(101) (a=b=3.776\AA, c=9.486\AA) and perovskite (a=b=c=5.87\AA) emphasize the differences in their structures, and we do not observe significant density fluctuations of ions near the interface before crystallization. The impact of VW crystallization on the morphological evolution of perovskite is shown in Figure \ref{fig:second}G, where the convergence of two growth fronts leads to the formation of a highly defected buried interface and a grain boundary. Through subsequent analysis of the buried region and grain boundary, we observe that these regions exhibit dangling bonds and incomplete formation of Pb-Br octahedra. Interestingly, Pb-Br edge-sharing octahedra stacking faults~\cite{chen_stabilizing_2021} and dislocations have crystallized in these regions, providing insights into the structural evolution of perovskite's buried interface.
\bigskip

\begin{figure*}
  \includegraphics[width=\linewidth]{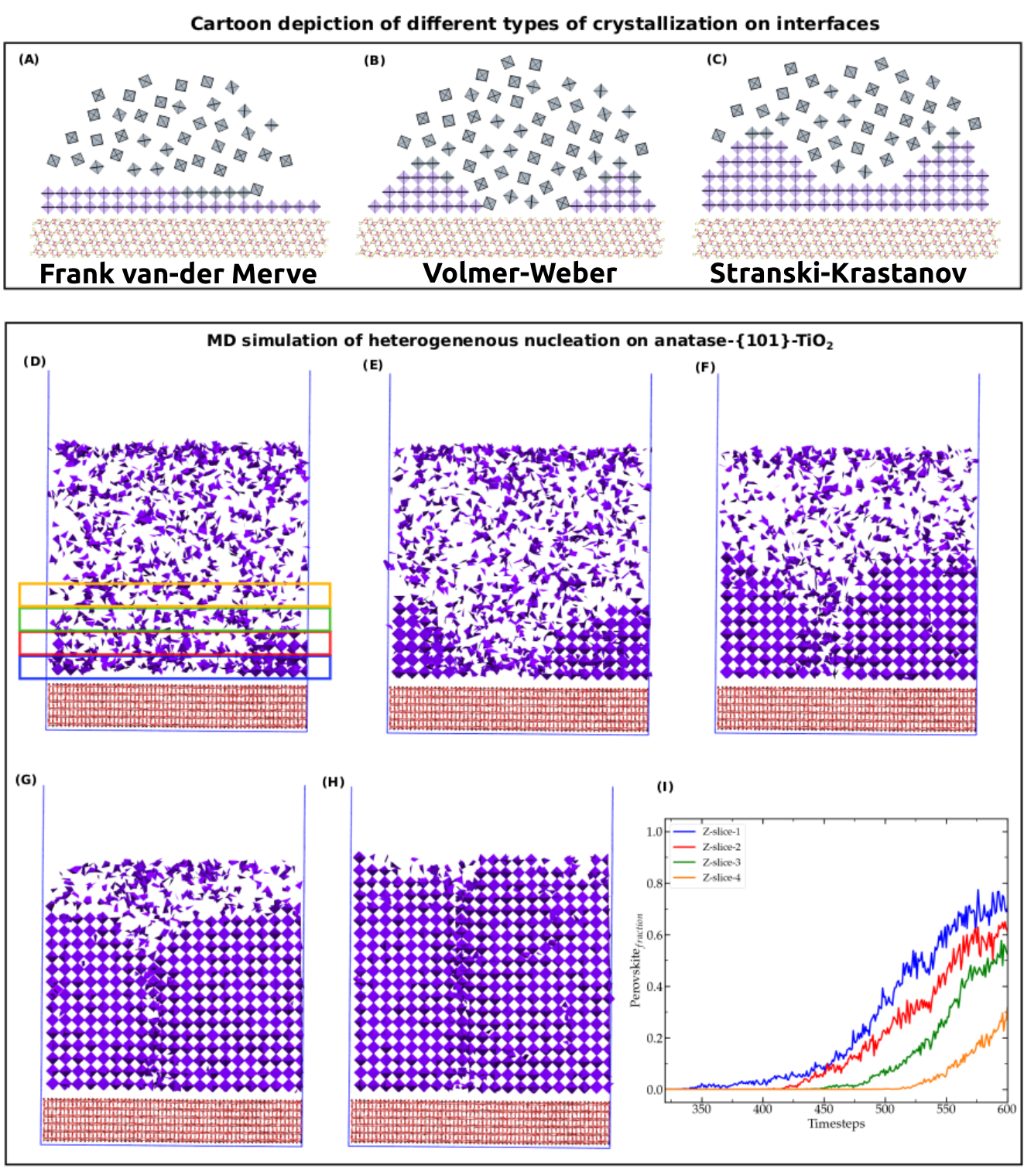}
  \caption{\textbf{Heterogeneous nucleation of a perovskite:} (A), (B), and (C) present cartoon depictions of commonly identified crystallization mechanisms. Panels (D) to (H) provide a visual representation of the typical evolution of heterogeneous nucleation on the interface, with only violet-colored Pb-Br octahedra shown for clarity. Figure (I) illustrates the temporal changes in the amount of perovskite in different layers calculated using L-SOAP-CV, as highlighted by the boxed region in Figure (D). All images were generated using VMD~\cite{humphrey_vmd:_1996}.}
  \label{fig:second}
\end{figure*}

\begin{figure*}
  \includegraphics[width=1.0\linewidth]{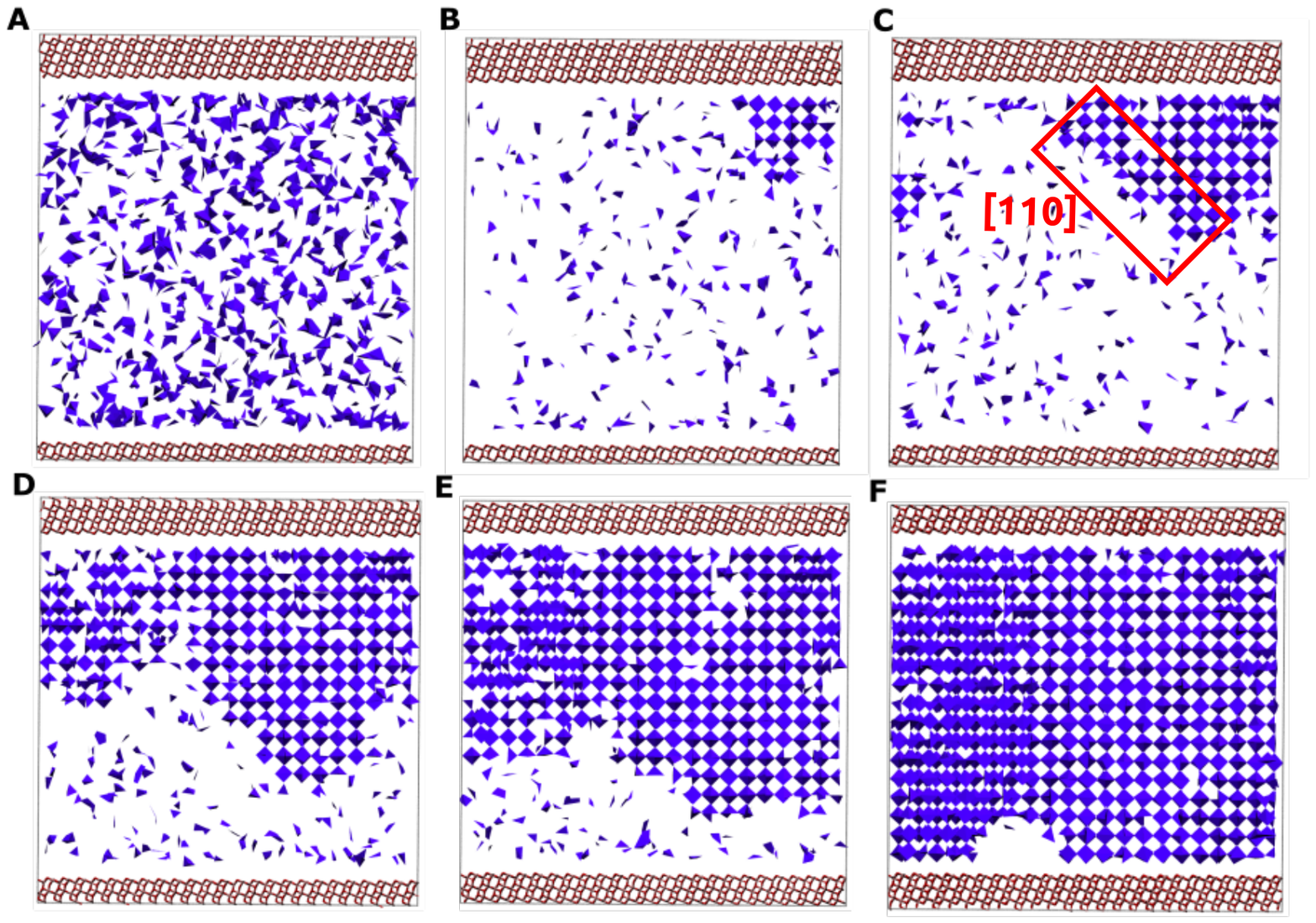}
  \caption{\textbf{Heterogeneous nucleation of a perovskite:} Figures A to F illustrate the typical progression of perovskite crystallization between flat TiO\textsubscript{2} substrates. For clarity, only Pb-Br octahedra are depicted in blue to facilitate the tracking and visualization of the crystallization process. The TiO\textsubscript{2} substrate is represented by a red-colored Ti-O bonded network. Images are generated with VMD software~\cite{humphrey_vmd:_1996}.}
  \label{fig:third}
\end{figure*}

Following setup (B), where a homogeneous mixture of ions is confined between flat TiO\textsubscript{2} interfaces, we conducted isothermal-isobaric simulations with moderate coolings for this system. Figure \ref{fig:third} provides a comprehensive view of the crystallization process, ranging from \ref{fig:third}A, depicting the initial configuration, to \ref{fig:third}F, representing the final crystallized state. Visual examination and atomic-level view provided in Supplementary Movie SM4 of these configurations, confirms the evident crystallization of perovskite on the substrate, as observed in Figure \ref{fig:third}B. Once nucleated, the growth proceeds rapidly, with the growth front reaching the other interface before nucleation occurs on the opposite side. Similar to the previous case, we observe a VW mechanism of growth, leading to the formation of perovskite islands, as depicted in Figure \ref{fig:third}C. Consistent with the previous system, we also observe the formation of grain boundaries and defective buried interfaces. However, a novel observation emerges in this case when the growth front reaches the other side. Notably, the formation of a nano-void is highlighted in Figure \ref{fig:third}F. Upon investigation, we again find that the presence of non-stoichiometric edge-sharing octahedra at the grain boundary and buried interface leads to a depletion of ions available for forming complete corner-sharing octahedra. Consequently, the formation of nano-voids is attributed to the emergence of non-stoichiometric structures resulting from ion consumption in the VW crystallization.

\bigskip

In both scenarios, the evolution of the perovskite microstructure reveals heteroepitaxial crystallization, accompanied by the emergence of grain boundaries, stacking faults, dislocations, and nano-voids. Earlier spectroscopic experimental investigations~\cite{dzhigaev_three-dimensional_2021, kosar_unraveling_2021, song_direct_2022, orr_imaging_2023, macpherson_local_2022} have substantiated that grain boundaries, dislocations, and stacking faults are among the most detrimental defects and substantial contributors to the degradation of halide perovskites. Recent experiments have further highlighted the rapid degradation of PSCs in the presence of nano-voids in the buried interface. Therefore, the elimination of these defects is crucial for achieving a stable perovskite morphology and advancing the industrialization of PSCs. From the above mentioned results, now a fundamental question arises: how can the observed defects be eliminated?

\subsubsection*{Lattice Matched Interface}

\begin{figure*}
\centering
  \includegraphics[width=1.00\linewidth]{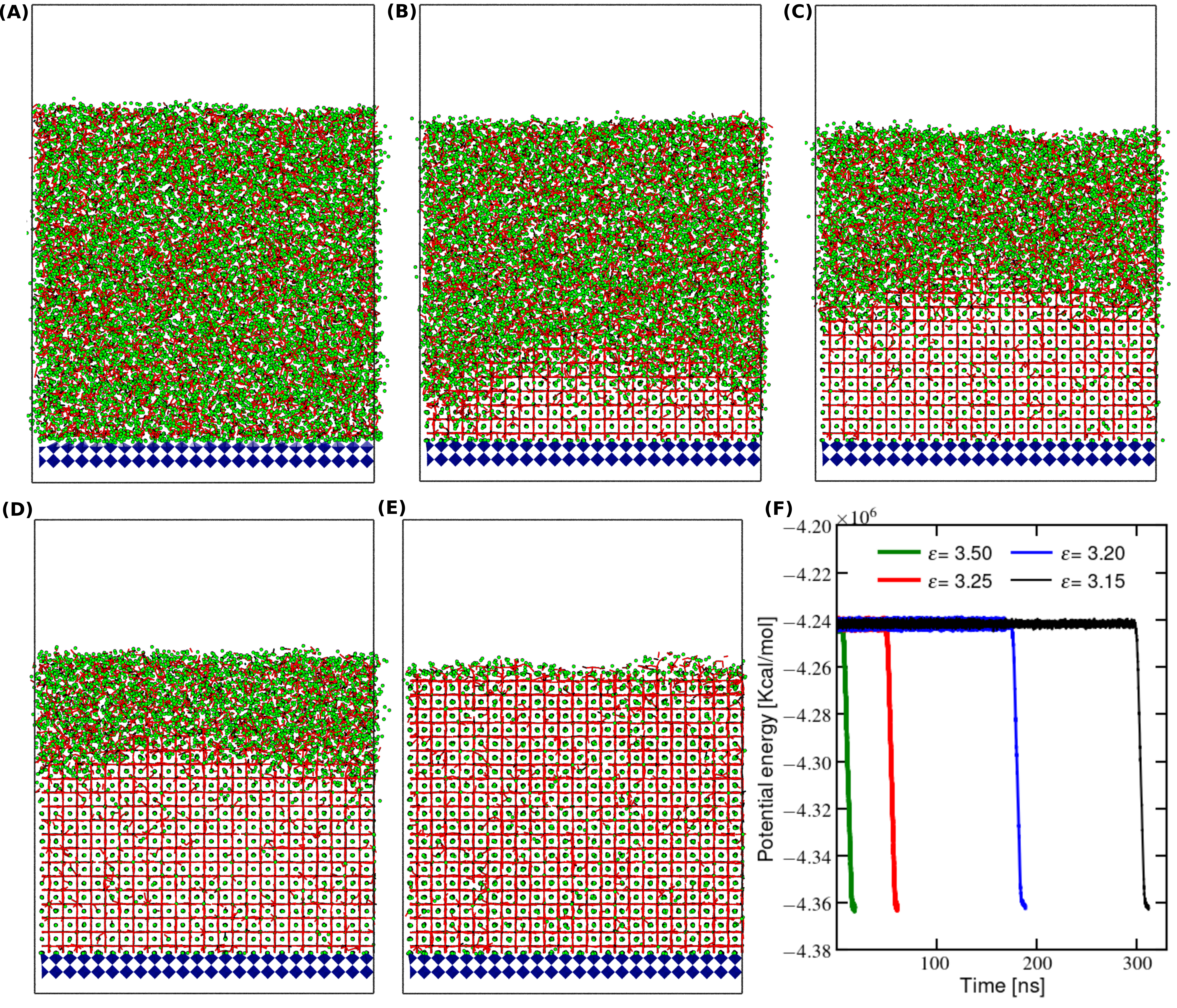}
  \caption{\textbf{Lattice matched interface:} (A) to (E) show the evolution of crystallization on an artificial lattice-matched interface; (F) displays the time evolution of the potential energy with varying ion-substrate interaction strength.}
  \label{fig:fourth}
\end{figure*}

In pursuit of this objective, we draw inspiration from previous research work on the crystallization and development of semiconductor heterostructures \cite{kroemer_polar--nonpolar_1987, alferov_history_1996} and blue LEDs~\cite{shuji_nakamura_gan_1991, amano_metalorganic_1986, oliver_growth_2005} and conduct MD simulations to study the lattice-matched perovskite crystallization. We construct an artificial lattice-matched substrate made of particles interacting by simplified Lennard-Jones (LJ) potential. An initial system is prepared by placing a homogeneous mixture of ions on this artificial perovskite-type LJ lattice (see Figure \ref{fig:fourth}A). Subsequently, brute-force MD simulations are carried out at moderate coolings, and the time evolution of the crystallization process is shown in Figures \ref{fig:fourth}A to \ref{fig:fourth}D. The nucleation begins by forming a uniform layer of perovskite at the blue-colored interface, as shown in Figures \ref{fig:fourth}B and in Supplementary Movie SM5. Unlike in the previous systems which followed the VW mechanism, the lattice-matched interface allows for the crystallization process to resemble the Stranski–Krastanov (SK) growth model (see Figure \ref{fig:second}C) where a uniform layer crystallizes on the substrate first, followed by the growth of islands on that layer, as partially evident in Figure \ref{fig:fourth}B. 

\begin{figure*}
\centering
  \includegraphics[width=1.00\linewidth]{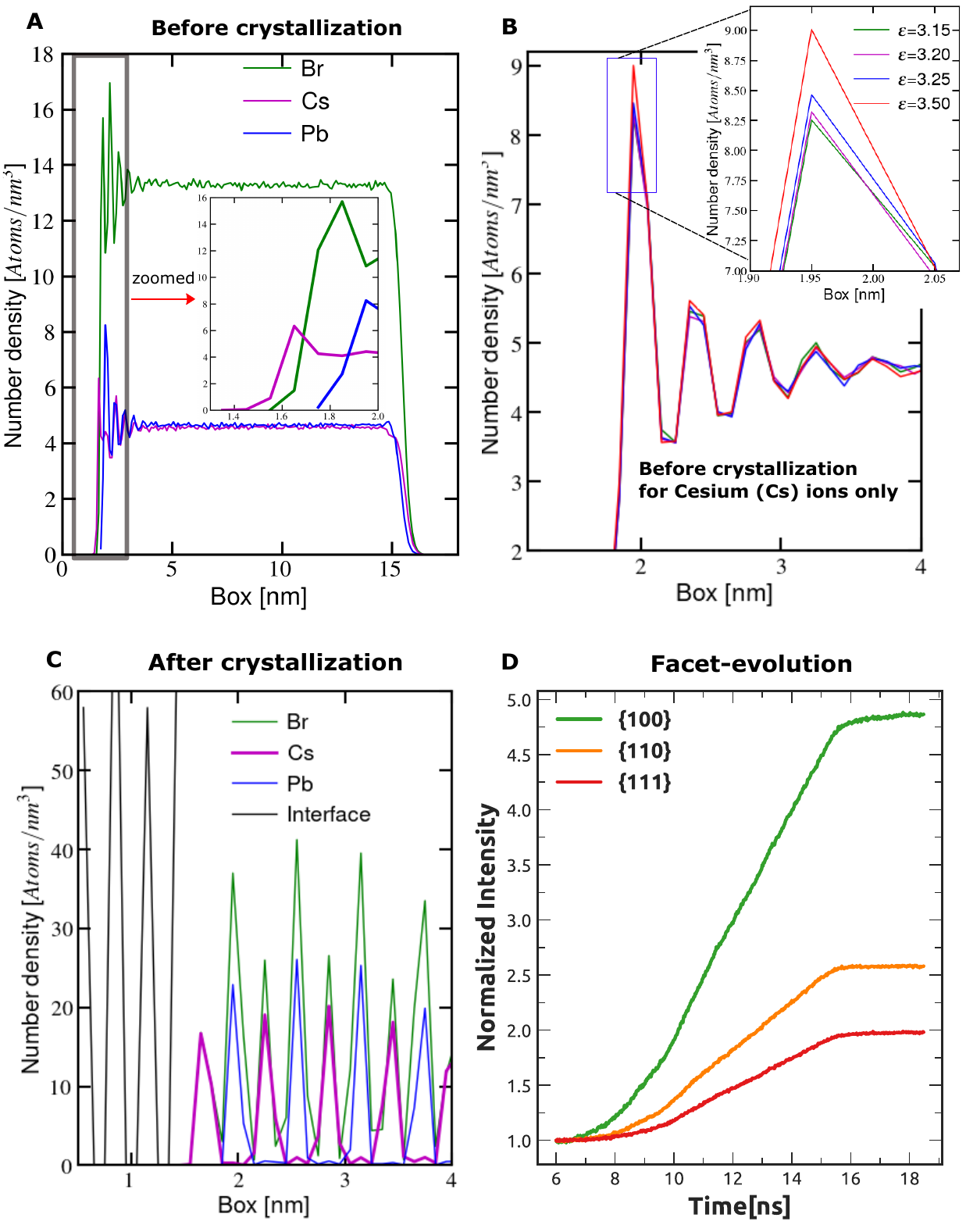}
  \caption{\textbf{Lattice matched interface:} (A), (B) and (C) represent the density profiles, while (D) illustrates the time evolution of corresponding X-ray peaks using the local Debye structure factor during crystallization.}
  \label{fig:fifth}
\end{figure*}

\bigskip

Previous experiments~\cite{song_evidence_2014} with different heterostructures have indicated that the Stranski-Krastanov (SK) type crystallization often involves density fluctuations at the interface, which help in forming a uniform crystalline layer. To explore this further, we examine the density of ions before and after crystallization, as depicted in Figure \ref{fig:fifth}A. We observe a higher concentration of ions near the substrate, particularly Cs ions closest to it. This trend becomes more pronounced after perovskite crystallization, as shown in Figure \ref{fig:fifth}C, where a complete layer of Cs ions forms on the substrate. This suggests that Cs ions may play a crucial role in promoting the uniform nucleation of the perovskite structure. To delve deeper into this phenomenon, we conduct additional simulations where we enhance the interaction strength between the substrate and the ions. Figure \ref{fig:fifth}B shows that the number of Cs ions near the interface increases with stronger interaction. This effect is further emphasized in Figure \ref{fig:fourth}F, where we observe an early decrease in potential energy, indicating the initiation of perovskite crystallization. To further validate these findings, we repeated the simulations multiple times and consistently observed similar outcomes. These results suggest that manipulating the interaction strength between the substrate and ions could potentially influence the nucleation and growth behavior of perovskite crystals, offering insights for controlling the crystallization process in future applications.

\bigskip

Finally, we investigate the differences in the bulk morphologies of perovskite crystals during lattice mismatch and lattice-matched crystallization processes. As highlighted in Figure \ref{fig:fourth}(C), the Volmer-Webber dominated crystallization process predominantly exhibits [110]-faceted perovskite crystals. In simulations involving lattice-matched substrates, we examine the localized facet evolution by employing local structure factors within a defined cutoff of crystallized atoms. We monitor the time evolution of the averaged intensities of X-ray peaks assigned to corresponding [100], [110], and [111] facets in the perovskite structure. While our periodic simulations are too limited in size to predict macroscopic crystal shapes accurately, the analysis of local structure evolution offers valuable insights into facet preferences for atom adsorption, providing glimpses into the final crystal morphology. In particular, Figure \ref{fig:fourth}(D) highlights that the crystal growth in lattice matching simulations is predominantly driven towards the [100]-faceted perovskite crystals direction.

\bigskip

\section{Discussion and Conclusion}\label{sec4}
Understanding and controlling the crystallization of halide perovskites represents a significant challenge for their industrialization. In this study, we try to address this challenge by building from fundamental quantum chemistry calculations of ionic interactions, followed by elementary all-atom MD simulations of perovskites on interfaces. We successfully observe the direct homogeneous and heterogeneous nucleation of halide perovskites. Through our investigation, we uncovered the formation of buried dislocations, grain boundaries, and nano-voids. A key observation is that defect formation is favored in mismatched heteroepitaxial systems, while lattice-matched substrates exhibit a uniform layered growth mode.

\bigskip

A pivotal insight from our simulations also lies in the potential to induce perovskite crystallization by modulating the interaction of ions with the interface. Consequently, perovskites form on the substrate covered by a thin wetting layer. Introducing a buffer layer could be instrumental in reducing the interfacial free energy, thereby transforming heteroepitaxial crystallization into epitaxial growth and minimizing the density of defects in the buried interface. Recent experiments~\cite{zhang_minimizing_2023, azmi_double-side_2024, gao_homogeneous_2024, xu_multifunctional_2024} have explored the use of various molecular additives to control the crystallization process and eliminate buried interfaces. However, the additives employed in halide perovskites are often selected through trial-and-error strategies. Further simulations and experiments would offer deeper insights into the impact of these additives on crystallization and help in the selection of specific types of molecular additives at interfaces. Moreover, recent experiments~\cite{dong_allinorganic_2022, ma_facet_2022} have also established that [100] facets exhibit superior electronic properties and stability compared to [110] and polar [111] surfaces. 

\bigskip

While our primary study provides valuable insights, it is essential to acknowledge its inherent limitations. First, inter-atomic potential we developed for CsPbBr\textsubscript{3} heterostructure simulations is qualitative and lacks the inclusion of polarization effects and other perovskite polymorphs. Additionally, our simulations are performed with a simplified setup compared to experimental conditions, excluding the simulation of solution mixtures and the effects of various solvents and additives on heterogeneous nucleation. Furthermore, we did not account for the effects of different types and facets of oxide substrates along with the presence of molecular additives. In future research, we aim to address these limitations by developing more accurate MLIPs for perovskites and commonly employed substrates, accounting for polarization effects and simulating various large-scale heterostructures, including those involving solutions. We also plan to investigate the impact of commonly used oxide substrates, such as SnO\textsubscript{2}, NiO, and Al\textsubscript{2}O\textsubscript{3}, on the buried interface. Additionally, we aim to explore oxide perovskite substrates, such as BaTiO\textsubscript{3}, which may demonstrate favorable lattice matching with halide perovskites.

\bigskip

With new developments in quantum chemical methodologies~\cite{filip_multireference_2019, nagy_approaching_2019, zhang_coupled_2019, booth_towards_2013, katukuri_ab_2022,  cui_systematic_2022, filip_hybrid_2023, zhao_rapidly_2024} and large scale MLIPs~\cite{fan_efficient_2017, tholke_torchmd-net:_2022, musaelian_learning_2023, gong_bamboo:_2024}, our computational framework presented in this study offers promising opportunities for investigating complex structural transformations across a diverse range of multi-species perovskite materials, for example, perovskite superconductors~\cite{bednorz_perovskite-type_1988, cava_superconductivity_1988, li_coexistence_2011, he_phase_2013}, multiferroics~\cite{wang_epitaxial_2003}, Josephson-junctions~\cite{hilgenkamp_grain_2002}, essential battery components~\cite{li_fluorinedoped_2016, lai_anti_2017, li_perovskite_2018, xu_high-performance_2019}, fuel cells~\cite{huang_double_2006}, and spintronics~\cite{ashoka_local_2023}. Achieving precise morphological control over heterostructures is crucial for advancing these materials toward industrial-scale applications.

\section{Methods}
\subsection{Inter-atomic potential}
We take inspiration from earlier studies~\cite{blazquez_madrid-2019_2022, zeron_force_2019} on ionic species and construct a reduced point charge inter-atomic potential for CsPbBr\textsubscript{3}. To fit the remaining Pb parameters with the existing species of Cs and Br, we employ the Complete Active Space Self Consistent Field (CASSCF) method~\cite{Roos1980,Siegbahn_1980,Siegbahn1981} to obtain binding curves for dimer fragments of the CsPbBr lattice, see Supplementary Figure 2. In this approach, one starts from a Hartree-Fock (HF)~\cite{Hartree1928,Fock1930,Roothan1951,Roothan1960} reference determinant $\ket{\Phi}$, built from a set of parametrized molecular orbitals $\{\boldsymbol{\phi}(\boldsymbol{\alpha})\}$, optimized to minimize the single-determinant energy,

\begin{equation}
    \widetilde{E} = \frac{\bra{\Phi}\hat H\ket{\Phi}}{\braket{\Phi}},
\end{equation}
where $\hat H$ is the system Hamiltonian. Disjoint sets of core (always filled), virtual (always empty), and active orbitals are defined. The space of all Slater determinants obtained by rearranging the active electrons among the active orbitals while preserving the spin and point group symmetry of the original HF reference is known as a complete active space (CAS) and customary labeled as ($n_\mathrm{elec}$,$n_\mathrm{orb}$), where $n_\mathrm{elec}$ is the number of active electrons and $n_\mathrm{orb}$ is the number of active spatial orbitals. The CASSCF wavefunction form is given by a linear combination of all determinants belonging to the CAS,

\begin{equation}
    \ket{\Psi(\{\mathbf{c}_\mu\},\{\boldsymbol{\alpha}\})} = \sum_{\mu \in \mathrm{CAS}} c_\mu\ket{\mu(\{\boldsymbol{\alpha}\})},
\end{equation}
where $\ket{\mu}$ denotes a Slater determinant. Both the parameters $\{\mathbf{c}_\mu\}$ and the underlying molecular orbital parameters $\{\boldsymbol{\alpha}\}$ are then optimized simultaneously in a self-consistent procedure to minimize the resulting energy,
\begin{equation}
    \widetilde{E} = \frac{\bra{\Psi}\hat H\ket{\Psi}}{\braket{\Psi}},
\end{equation}

By allowing for orbital relaxation from the HF reference and the inclusion of multiple determinants of comparable weights, the CASSCF approach excels at capturing strong or static correlation effects and therefore provides good descriptions for dissociation binding curves. We run first-principle CASSCF calculations using an (8,8) CAS for closed-shell and a (7,8) CAS for open-shell species in the def2-TZVPP basis set~\cite{leininger1996a, weigend2005a, metz2000a}, with corresponding effective core potentials, using the PySCF program~\cite{pyscf} and extract the inter-atomic potential parameters listed in the Supplementary Information. Using these parameters, we find that the perovskite structure is stable in its original form at finite temperatures. We test this inter-atomic potential against finite temperature experimental lattice parameters and melting point by co-existing simulations illustrated in Supplementary Figure 2. 

\bigskip

The CASSCF approach is very well-suited to study the strong correlation effects present in dissociation processes, but it scales exponentially with the size of the active space considered, making it challenging to expand to larger systems. Therefore, if we wish to expand this approach to more electrons or larger fragments, other methods may be necessary. Approaches based on Monte Carlo simulations of the wavefunction in Hilbert space have been shown to be effective in treating this type of problem, at reduced computational cost~\cite{filip_multireference_2019,dobrautz_spin-pure_2021,weser_stochastic_2022,filip_hybrid_2023,zhao_rapidly_2024}, while allowing for flexible definitions of the active space that may be optimized for the interaction in question.

\subsection{Machine learning potential}
The NequIP potentials~\cite{e3nn, batzner_e3-equivariant_2022, e3nn_paper} employed in this study are derived from \textit{ab-initio} molecular dynamics (AIMD) simulations based on DFT. The training methodology involves the utilization of two interaction layers with a radial cutoff of 8\AA, incorporating interatomic distances encoded within a basis comprising eight Bessel functions. Feature representations are selected up to rotation orders of L=2. The loss function incorporates forces, energies, and stresses obtained from DFT calculations. Optimization is performed using the Adam optimizer with a learning rate of 0.005 and a batch size of 10. The training process is executed on a V100 GPU.

\subsection{DFT calculations}
In this study, all DFT calculations are conducted employing the r\textsuperscript{2}SCAN+rVV10 functional~\cite{ning_workhorse_2022}, with a 600 eV plane-wave-basis cutoff, PBE potentials, and the projector augmented wave (PAW) formalism for electronic minimization, implemented within the VASP (Vienna ab initio simulation package) code~\cite{kresse_ab_1993, kresse_efficiency_1996}. To generate the training dataset, ab initio molecular dynamics (AIMD) simulations are carried out for 135-atom supercells, corresponding to a 3x3x3 unit cell of cubic-CsPbBr\textsubscript{3}, including both crystalline and molten CsPbBr\textsubscript{3} phases at temperatures ranging from 600 K to 1000 K, with intervals of 100 K, for a duration of 2 ps. These simulations are performed under the isothermal-isobaric (NPT) ensemble, utilizing a time step of 2 fs. The pressure and temperature are maintained using the Parrinello-Rahman barostat~\cite{parrinello_polymorphic_1981} and the Langevin thermostat, respectively. For these relatively large supercells, only Gamma point sampling in the Brillouin zone is employed. The melted configuration is obtained by heating the crystal to a very high temperature of 2300 K.

\subsection{Landmark-SOAP Collective Variable (L-SOAP-CV)}
Crystallization in multi-species materials is a complex phenomenon characterized by the emergence of a crystalline nucleus through the local ordering of diverse chemical species. Various CVs, often tailored to specific systems or phases, have been introduced to quantify and scrutinize the intricacies of the crystallization process. Inspired by prior investigations on phase transitions, we draw insights from studies on bond-orientational order parameters~\cite{steinhardt_bond-orientational_1983}, accurate particle-based measures~\cite{lechner_accurate_2008}, and morphology-aided order parameters~\cite{yu_order-parameter-aided_2014, chen_morphology_2021, rogal_reaction_2021}. Taking inspiration from these foundations, we formulate simplified one-dimensional CVs and hereafter, we provide details on the \textbf{landmark-SOAP} CV and its application in the present study. In essence, we propose applying the SOAP kernel~\cite{Bartok2013} to measure the similarity between hand-picked landmark atomic environments, effectively defining the target local nucleated states, and atomic structures, e.g. extracted from nucleation simulations. The SOAP representation~\cite{Bartok2013} characterizes the neighborhood of an atom $i$ within an atomic structure $X$ by extracting invariant $n$-body correlations from the atomic density field
\begin{equation}
    \rho(\mathbf{r}, a; X_i, r_c, \sigma) = \sum_{j \in X_i} \delta_{a a_j} \exp\left[\frac{(\mathbf{r}-\mathbf{r}_{ij})^2}{2 \sigma^2}\right] f_c(r_{ij}),
\end{equation}
where the sum $\sum_{j \in X_i}$ runs over the atoms $j$ of type $a_j$ (including atom $i$) that belong the neighborhood of atom $i$, $\mathbf{r}_{ij}=\mathbf{r}_j-\mathbf{r}_i$, $r_{ij}=\|\mathbf{r}_{ij}\|$, $f_c$ is a smooth cutoff function with cutoff radius $r_c$, and $\sigma$ is a smoothing parameter.
For simplicity, we focus on the SOAP powerspectrum expanded on a basis of radial functions and spherical harmonics given by
\begin{equation}
    p^{(i)}_{a_1 n_1 a_2 n_2 l} = \frac{1}{\sqrt{2l+1}}\sum_m (-1)^m c^{(i)}_{a_1 n_1 l m} c^{(i)}_{a_2 n_2 l (-m)},
\end{equation}
where $c^{(i)}_{a n l m}$ are the expansion coefficients of the density field (see Ref.~\cite{Musil2021} for the full expression), and the indices $n$ and ($l$,$m$) refer to the radial basis and spherical harmonics.
The SOAP powerspectrum coefficients are invariant with respect to atom permutations, global translations, global rotations, and reflections of the atomic coordinates.
Moreover, they encode the 3-body correlations within the environment of atom $i$ with the different atomic identities present in the system. Therefore, this representation is able to differentiate with great accuracy a wide range of atomic environments.~\cite{Musil2017, De2017} To transform the characterization of atomic neighborhoods given by the SOAP representations into a CV, we use the SOAP kernel~\cite{Bartok2013} 
\begin{equation}
    k_{\zeta}(L_i, X_j) = \left[\frac{\mathbf{p}^{(L_i)} \cdot \mathbf{p}^{(X_j)}}{\|\mathbf{p}^{(L_i)}\| \|\mathbf{p}^{(X_j)}\|} \right]^{\zeta},
\end{equation}
to measure the similarity of the environment of atom $L_i$ and atom $X_j$ through the lens of the SOAP powerspectrum. Note that this similarity measure has the same properties~\cite{Musil2021rev} (invariances, information content, \dots) as the representation it is based on, e.g. two configurations that are a reflection of each other look identical with this metric. To make this CV more concrete, we take the example of the nucleation of perovskite. We measure the similarity between landmark environments $\{L_i\}_{i=1,\dots,R}$ extracted from the perfect perovskite crystal, e.g. centering on the Pb atoms, with comparable environments extracted from nucleation simulations, we can estimate the fraction of the material $X$ that has nucleated
\begin{equation}
N_c (X) =  \sum_j \sigma_{m,k} \left( \mathrm{softmax} \left[ k_{\zeta}(L_1, X_j), \dots, k_{\zeta}(L_R, X_j) \right] \right)
\label{eq:seed_count}
\end{equation} 

where $\sigma_{m,k}(x)= \frac{1}{2} \left( \mathrm{tanh}[\frac{k(2x'-1)}{2\sqrt{x'-x'^2}}] + 1\right)$ with $x'=x-m+0.5$ is a switching function, and the parameters $m$ and $k$ control respectively the position and the steepness of the switch. Typical values are $m=0.97 $ and $k=200$. We note here that the versatility of the SOAP representation~\cite{Bartok2013, Musil2021rev} ensures that the landmark-SOAP CV can be applied to a wide range of materials while the use of landmark environments allows for tuning this CV to resolve the differences between many condensed polymorphs. 

\subsection{Details of MD simulations with LAMMPS}
Before performing production runs, we achieve system equilibration through 10 ns isothermal-isobaric simulations conducted above the melting temperature. Our simulations employ a 1.0 nm cutoff for nonbonded interactions, particle-mesh Ewald for handling electrostatic interactions, a velocity re-scaling thermostat~\cite{bussi_canonical_2007} with a relaxation time of 0.1 ps for temperature control, and a Parrinello-Rahman barostat~\cite{parrinello_polymorphic_1981} for pressure control, featuring a relaxation time of 10 ps. The classical molecular dynamics simulations are executed using the Large-Scale Atomic/Molecular Massively Parallel Simulator (LAMMPS) code (version 31 Mar 2017)~\cite{thompson_lammps_2022}.

\subsubsection{Co-existing simulations}
The initial structures as illustrated in Figure \ref{fig:coexist} are constructed by combining seeded structures of CsPbBr\textsubscript{3} with the molten configuration. Seeds oriented along [100], [110] and [111] directions are generated using ASE utilities. Melted configurations are created either by melting crystalline seeds at elevated temperatures via inter-atomic potentials or by generating supercells directly from molten structures through AIMD simulations. The final configuration involves an equal number of atoms in seed and melt: 768 f.u. for scaled charge based simulations and 128 f.u. for MLIPs. The MLIP-based simulations are conducted using the $pair{\textunderscore}nequip$ interface with LAMMPS code.  

\subsubsection{Crystallization on interfaces}
The simulation systems shown in Figures \ref{fig:second} and \ref{fig:third} are prepared by depositing homogeneous mixtures of ions on flat (101)-anatase-TiO\textsubscript{2} substrate. A point charge inter-atomic potential~\cite{bandura_derivation_2003, matsui_molecular_1991} was used for TiO\textsubscript{2}. During constant volume simulations, substrates are kept frozen. For lattice-matched simulations in Figure \ref{fig:fourth}, we prepared a larger simplified perovskite lattice with a similar lattice constant as CsPbBr\textsubscript{3} and particle-particle inter-actions are described by Lennard-Jones (LJ) potential. Lattice matched simulations consist of approximately 60000 atoms. 

\backmatter

\bmhead{Supporting Materials}
\begin{itemize}
    \item Perovskite-simulations.ipynb: A python-notebook to perform MD simulations of perovskites 
    \item Quantum-chemistry.ipynb: A python-notebook to perform CASSCF calculations in PySCF
    \item SOAP-analysis.ipynb: A python-notebook to perform SOAP calculations 
    \item All simulations data-set including trajectories: 10.5281/zenodo.10975237
\end{itemize}

\bmhead{Acknowledgements}
This research is funded by Swiss National Science Foundation (SNSF) Postdoc.Mobility fellowship P500PN\_206693. F.M. acknowledges support from the SNSF under the Postdoc.Mobility fellowship P500PT\_203124, and from the Physics department and Freie Universität Berlin for computing time. M.A.F. is grateful to Peterhouse, University of Cambridge for support through a Research Fellowship. P.A. is indebted and greatly thankful to Yusuf Hamied Department of Chemistry for hosting and providing necessary computational resources through Rogue-GPU cluster, Nest-CPU cluster, UK Materials and Molecular Modelling Hub YOUNG CPU/GPU cluster, which is partially funded by EPSRC (EP/T022213/1, EP/W032260/1 and EP/P020194/1). This work used the ARCHER2 UK National Supercomputing Service (https://www.archer2.ac.uk). P.A. is immensely thankful to Professor Daan Frenkel for several key discussions, large dose of encouragements and key guidance that led to fruition of this research. P.A. is sincerely thankful to Professors Michael Graetzel, Angelos Michaeledis, Mike Payne, Alex Thom, Ali Alavi, Carlos Vega, Peter Bolhuis, Rob Jack, Alessandro Laio, Rachel Oliver, Erin Johnson and Michiel Sprik for highly inspiring and insightful discussions.

\bmhead{Contributions} 
P.A. conceived, formulated and conceptualized research. M.A.F. designed and performed CASSCF calculations. F.M. created L-SOAP-CV and designed an interface with PLUMED2 software. P.A. designed both classical and machine learning interatomic potentials, designed and performed all classical, MLIPs and \textit{ab-initio} MD simulations, and conducted analysis. C.C. read the manuscript and provided some comments. P.A., M.A.F and F.M. discussed research and wrote the manuscript together.

\bmhead{Declarations}
Authors declare no conflicting interest.

\bibliography{sn-bibliography}

\end{document}